\documentclass[12pt]{article}

\def\mytitle{On the Equivalence of Phase and Field Charges}

\usepackage{amssymb,bbm}  
\def\setZ{\mathbbm Z}     
\def\setS{\mathbb S}      

\def\be{\begin{equation}}
\def\ee{\end{equation}}
\def\ket#1{ | #1 \rangle }                  
\def\braket#1#2{ \langle #1 | #2 \rangle }  
\def\Hil{{\cal H}}                          

\renewcommand{\baselinestretch}{1.1}
\begin{document}
\sloppy

\thispagestyle{empty}

\vspace*{-20mm}


\vspace{20mm}

\noindent{\Large\bf \mytitle}

\vspace{13mm}

\hfill\parbox{13cm}{

{\large Holger Lyre}

\vspace{8mm}

{\small
Philosophy Department,
University of Bonn,\\
Am Hof 1, D-53113 Bonn, Germany,
E-mail: lyre@uni-bonn.de
}

\vspace{9mm}

March 2003

\vspace{9mm}

\renewcommand{\baselinestretch}{.99}\large\small

{\bf Abstract.} The analysis of the gauge principle as a mere
passive symmetry requirement leads to the conclusion that the
connection term in the covariant derivative is flat and that local
phase transformations are without any empirical significance in
analogy to coordinate transformations. Nevertheless, the 
Aharonov-Bohm effect shows the physical significance of 
the non-trivial holonomy of a flat connection. On this basis
the proposal of a new kind of charge, the phase charge, is made,
understood as the coupling strength of the particle to the holonomy.
The equivalence of phase and usual field charge must be tested 
experimentally in terms of an Aharonov-Bohm effect with muons 
or tauons, for instance.

\renewcommand{\baselinestretch}{1}\large\normalsize

}\hspace*{5mm}

\vspace{13mm}


\section{Gauge principle: received and passive view}
\label{passive}

The received view of gauge theories and their underlying gauge principle
is that we are able to derive the coupling of a formerly free field to
an interaction field from the requirement of local gauge covariance.
The usual derivative $\partial_\mu$ must be replaced by a covariant
derivative 
\be
\label{cov-der1}
D_\mu = \partial_\mu - iq A_\mu(x),
\ee
where $A_\mu$ is a Lie~algebra-valued 1-form---mathematically 
a connection in a principal fiber bundle, physically the gauge potential
of an interaction field. However, this gauge argument would be
close to a miracle, if the connection, which arises from the local
symmetry requirement, were non-flat (i.e.\ with non-vanishing curvature).

But this, of course, is not---and can hardly be---the case.
What rather happens is that we explicitly allow for a freedom
in the choice of the position representation of the wave function
(figuring as a free Schr\"odinger or Dirac field
in the matter field equation from which the gauge principle starts).
To see this immediately, consider a system of vectors
$\Big\{\ket{\phi}\Big\}$ spanning an abstract Hilbert space $\Hil$,
such that a wave function in the position representation $\ket{x}$
simply reads $\Psi(x) = \braket{x}{\phi}$.
Accordingly, local gauge transformations are expressed as
$\ket{x'} = e^{i \chi(x)} \ket{x} = \hat U \ket{x}$.
Such a change of representation must apply as well to all operators
$\hat O$ acting on $\Hil$, viz.\ $\hat O' = \hat U \hat O \hat U^+$.
In particular, for the derivative operator we get
\be
\label{cov-der2}
D_\mu = \partial_\mu - i \partial_\mu \chi(x).
\ee
Identifying the gradient of the phase with the gauge potential $A_\mu$
(multiplied by some charge $q$ in order to get the units%
\footnote{If not otherwise stated we set $c=\hbar=1$ throughout this paper.}
straight)
\be
\label{chi-alpha}
\partial_\mu \chi(x) = q \ A_\mu(x)
\ee
leads to  (\ref{cov-der1}). Obviously, $A_\mu$ is a flat connection.

Thus, the celebrated gauge principle is not sufficient 
to `derive' the coupling to a new interaction-field,
but rather makes the built-in covariance under local gauge transformations
explicit. We are, from the mere logic of the argument, not enforced
to consider the connection term non-flat. 
Moreover, the Noether current connected to the covariance of the 
Dirac Lagrangian under global $U(1)$ transformations is just the
probability density current $S^\mu=\bar\psi \gamma^\mu \psi$ and
not the charge current $j^\mu=q \ S^\mu$, since there simply is no
charge occurring in the Dirac equation.
Again, in the standard textbook presentation the charge
is put in by hand by means of (\ref{chi-alpha}).
Call all this the \textit{passive view} on the gauge principle, since local
gauge transformations are treated in full analogy to coordinates---coordinates,
however, in the fiber bundle rather than in the space-time base space.
The analogy is complete if we draw a comparison to the Levi-Civita
connection in General Relativity. Here, Christoffel symbols already
occur in the geodesic equation simply because of curvilinear coordinates
in flat Minkowski space---without entailing a real gravitational
field, i.e.\ non-vanishing Riemann curvature.


\section{The non-observability of local phase transformations}

Let us now consider the gauge principle's local phase transformations
(a.k.a. gauge transformations of the first kind) in more detail.
Under such transformations the wave function yields
\be
\psi(x) \ \to \ \psi'(x) = \psi(x) e^{i \chi(x)}.
\ee
Thus, we obtain $\partial_\mu \psi(x) \to \partial_\mu \psi'(x) =
e^{i \chi(x)} \Big( \partial_\mu + i \partial_\mu \chi(x) \Big) \psi(x)$,
which confirms the covariant derivative (\ref{cov-der2}),
i.e.\ Dirac equations
$(i \partial_\mu \gamma^\mu - m) \psi(x) = 0$ and
$(i D_\mu \gamma^\mu - m) \psi'(x) = 0$ are equivalent.
We also get with $\hat{p}_\mu = - i \partial_\mu$ and
$e^{i \chi}=e^{i px}$ the phase transformation behavior
$\chi \to \chi - \partial_\mu \chi(x) \cdot x^\mu$,
which leads to the holonomy
\be
\label{holonomy}
\Delta \chi = \oint \partial_\mu \chi(x) \ d x^\mu.
\ee
Of course, written as such and provided that spacetime is simply 
connected, expression (\ref{holonomy}) is trivial and $\Delta \chi=0$.
We will come back to this point in a moment.

As an intermediate step, let us ask for the possibility to observe local
phase transformations. A widespread argument says that---in contrast
to the passive view---the phase transformed
wave function leads to new expectation values.
We get, for instance, for the momentum operator
$\hat{p}_\mu \psi' = (p + \partial_\mu \chi ) \psi'$
as opposed to $\hat{p}_\mu \psi = p \psi$.
But this is of course misleading since we must use the transformed momentum 
operator $\hat{P}_\mu = \hat{p}_\mu - \partial_\mu \chi$ corresponding to 
(\ref{cov-der2}) and, thus, $\hat{P}_\mu \psi' = \hat{p} \psi'$.
Indeed, just like their global counterparts local phase transformations
do not change any expectation values at all.

Another argument can be found in 't Hooft (1980), 
who considers an ordinary double-split experiment showing
an interference pattern. Inserting a phase shifter behind the slit
in one of the two paths results in a corresponding shift of the
interference pattern---which can be calculated from (\ref{holonomy}).
't Hooft now argues that the phase shifter can be seen as a
realization of a local phase transformation. But this is impossible
since local phase transformations would then change the holonomy,
which is, however, invariant under (global and local) $U(1)$.
What is rather observed in this case is the relative phase shift.
Let $\psi = \psi_I + \psi_{II}$, where $\psi_I$ and $\psi_{II}$ are
the two partial wave functions on the two paths $I$ and $II$, then, 
say, a $\lambda/4$-phase shifter in path $I$ corresponds to
$\psi_I \to \psi_I e^{i \lambda/4}$ and hence
$\psi \to \tilde{\psi} = \psi_I e^{i \lambda/4} + \psi_{II}$,
where $\psi$ and $\tilde{\psi}$ obviously do have
different expectation values in general
(cf.\ Brading and Brown 2003). 
We must therefore very well distinguish between
relative phase shifts and local phase transformations.
The former are observable, the latter clearly are not.


\section{``Phase charges'' and Aharonov-Bohm effect}

As already mentioned, only non-trivial holonomies are of interest.
In this case the loop cannot be contracted to a point
and the underlying fiber bundle is non-trivial, too. The double-slit
experiment or the Mach-Zehnder interferometer are cases at hand.
Another example is provided by the Aharonov-Bohm (AB) effect.
Here we have an observable effect despite the fact
that the connection is flat.
At first glance, this seems to contradict the passive view statement
from Sect.~\ref{passive}, where we made the claim that from flat 
connections alone the physical situation can hardly be changed.
However, the ultimate cause of the AB effect is of course the magnetic field
confined to the region of the solenoid in the experimental setting.
Nevertheless, there is no magnetic field in the region of the electron
(the configuration space of the electron). Here the connection is flat,
but the non-trivial holonomy of this connection causes observable effects.
Thus, the lesson of the AB effect decidedly is that we must consider
the holonomy outside the solenoid as a real entity!
This conclusion is inevitable if we want to avoid an interpretation
that either considers gauge-dependent quantities as physically 
real---such as the gauge potential---or does not conform to the idea of
local action---which happens in the case where we allow for a non-local
interaction between the confined magnetic field and the electron
wave function (Eynck, Lyre, Rummell 2001). 

As a real entity, the holonomy couples to some property of the electron,
and this, in the usual picture, is just the charge $q$. We shall, however,
rather write $q^{(p)}$ (the superscript will become clear soon).
In the AB case there is, indeed, more to (\ref{chi-alpha}) than a mere rewriting
of the phase function, since from (\ref{holonomy}) together with (\ref{chi-alpha})
we get the observed phase shift as a function of $q^{(p)}$ and the magnetic flux
\be
\label{q_phase}
\Delta \chi
\stackrel{(\ref{chi-alpha})}{=}
q^{(p)} \oint A_\mu d x^\mu =
q^{(p)} \int F_{\mu\nu} d x^{\mu\nu} = q^{(p)} \Phi_{mag} .
\ee

But what really is the origin of the charge $q^{(p)}$ in (\ref{q_phase})?
Obviously, it is a certain property of the electron, but it is not
the property of a `usual' charge being the source and drain of the
electromagnetic field, since there is no field in the configuration
space of the electron (or, at least, we may abstract from it).
The AB effect itself has its origin in the topological nature of
the non-trivial holonomy, since mappings $\setS^1 \to \setS^1$ 
from the electron's configuration space to the gauge group
are non-trivial and constitute the fundamental group $\pi_1(U(1)) = \setZ$.
This holonomy now, as a physical entity, couples in some way to the electron,
so we may very well interpret $q^{(p)}$ as the \textit{coupling strength}
between the electron wave function and the holonomy.

Let us call the `usual' charge---the source and drain of the
field---the active or passive \textit{field charge} $q^{(f)}$---in 
full analogy to the active or passive gravitational mass
(the charge of the gravitational field).
By way of contrast, $q^{(p)}$ can be called the \textit{phase charge},
since it originates from the phase factor of the wave function only.
A first argument for this conceptual maneuver of distinguishing
two different kinds of charges is that we have in fact no \textit{a priori}
reason to identify them. A more compelling physical argument is
that $q^{(p)}$ is in principle measurable in a isolated experiment.
Whereas $q^{(f)}$ is tested whenever we perform measurements
where charges figure as sources or drains of the electromagnetic field
(e.g.\ in measuring the Coulomb force between two electrons),
$q^{(p)}$ only becomes visible, if we consider (\ref{holonomy})
and its observable consequences.
And this is exactly what happens in the AB effect.

All this leaves us with a remarkable conclusion:
\textit{It might very well be the case that particles
with one and the same field charge do have different phase charges
and, therefore, show different AB effects.}
This can clearly be seen from (\ref{q_phase}).
We have (on the l.h.s.) the observable shift of the interference
pattern, which is proportional to the phase charge $q^{(p)}$
and the magnetic flux $\Phi_{mag}$ (on the r.h.s.).
The latter can be measured independently by testing, for instance,
the Lorentz force of the magnetic field on electrons and muons
(which is obviously the same, because of their common field charge).
However, from the way (\ref{q_phase}) was derived,
we have no reason to expect the same interference shift
in an AB experiment for electrons as compared to muons.

Note that the claim is not that in an actual experiment
electrons and muons will show different AB effects.
One would, in fact, expect the same $q^{(p)}$ for both.
The above arguments are rather intended to show that
there is no theoretical principle in our known physics
which precludes the possibility of differing phase charges.
Insofar as $q^{(p)}$ and $q^{(f)}$ are conceptually different,
their equivalence
\be
\label{gep}
q^{(p)} = q^{(f)}
\ee
must be tested experimentally.
Therefore the real claim here is that some of our Standard Model's
experiments---involving topological effects from flat
connections---are in fact ``null experiments'' on the
\textit{equivalence principle of field and phase charge}.


\section{A gauge theoretic equivalence principle}

In Lyre (2000) 
the attempt was made to propose a gauge theoretic generalization
of the equivalence principle. The analogy is indeed striking:
Field charges and gravitational mass appear in the field equations
of a field theory (e.g.\ Maxwell or Einstein equations), whereas
phase charges and inertial mass appear in the corresponding
equations of motion (e.g.\ Dirac or geodesic equations).%
\footnote{The presentation in Lyre (2000) was insufficient,
since no isolated experiment for $q^{(p)}$ was proposed.
This is the main task of the present paper.
A rather rhetoric move is to rename the misleading term
`inertial charge' of the former paper into `phase charge' here.}
The generalized equivalence principle is then intended to fill
the explanatory gap in the architecture of a gauge theory
arising from the mere passive view of the gauge principle.
This gap is simply the following:
Let ${\cal L}_D = \bar\psi (i \partial_\mu \gamma^\mu - m) \psi$
be the Dirac Lagrangian and ${\cal L}_{coup}=j_\mu A^\mu$
the inhomogeneous `coupling' part of the Lagrangian
\be
\label{LagD}
{\cal L}_D' = {\cal L}_D + {\cal L}_{coup},
\ee
which arises due to the replacement of the usual derivative
by the covariant derivative (based on the requirement of local
gauge covariance). As we have seen, however, the covariant derivative
(\ref{cov-der2}) only entails a flat connection in ${\cal L}_{coup}$.
There is, on the other hand, the Maxwell theory with the Lagrangian
\be
\label{LagM}
{\cal L}_M' = {\cal L}_M + {\cal L}_{coup}.
\ee
Here ${\cal L}_M =  \frac{1}{4} F_{\mu\nu} F^{\mu\nu}$ is the kinetic term
of the free Maxwell field and ${\cal L}_{coup}$ the inhomogeneous term
including field charges.
In this case the connection in ${\cal L}_{coup}$ is non-flat.

In order to arrive at the full Lagrangian of the Dirac-Maxwell
gauge theory (or QED, analogously), we have to combine ${\cal L}_D'$
and ${\cal L}_M'$ in order to get
\be
\label{LagDM}
{\cal L}_{DM} = {\cal L}_D + {\cal L}_{coup} + {\cal L}_M.
\ee
But, obviously, ${\cal L}_{coup}$ figures in two different meanings here.
Because of its mere passive nature, the gauge `principle' does not
allow to generalize from a flat connection in (\ref{LagD})
to a non-flat connection in (\ref{LagM}).
Thus, the two ${\cal L}_{coup}$-terms cannot simply be identified.

There is, however, the possibility of non-trivial holonomies 
with a phase charge $q^{(p)}$ in the inhomogeneous term in (\ref{LagD}), 
which we therefore indicate as ${\cal L}_{coup}^{(p)}$ as opposed to 
${\cal L}_{coup}^{(f)}$ in (\ref{LagM}). 
From the equivalence of field and phase charges (\ref{gep}) we then get
\be
{\cal L}_{coup}^{(p)} = {\cal L}_{coup}^{(f)} \equiv {\cal L}_{coup}
\ee
and, thus, the desired Lagrangian (\ref{LagDM}).
The $U(1)$ gauge theory is therefore not based on the physically
vacuous gauge `principle', a mere passive symmetry requirement,
but on the gauge theoretic equivalence principle,
which in the form of (\ref{gep}) must be verified empirically.


\section{Discussion and conclusion}

The concept of a phase charge and, hence, the gauge theoretic
equivalence principle is based on the existence of a non-trivial holonomy.
Unfortunately, there are no AB effects for higher $SU(n)$ groups,
since $\pi_1(SU(n)) = 1$. As far as other topological effects in
gauge theories are concerned (e.g.\ instantons or $\theta$-vacua),
it is not so clear whether they do perhaps only arise
because of some clever approximations (e.g.\ the assumption of
\emph{vanishing fields} in the infinity of \emph{Euclidean spacetime}
for $SU(2)$-instantons). In these cases the holonomy
should not be considered a really existing entity.
This then means that our considerations only apply for $U(1)$
and that no simple extension of the gauge theoretic equivalence
principle to the general Yang-Mills case will be possible.

One should also mention that the analogy between the gauge theoretic
and the usual gravitational equivalence principle, striking as it may be,
is of course only a heuristic one. As already mentioned, it certainly
breaks down if we compare the concepts of phase charge and inertial mass.
The reason for this might very well be seen in the
classical nature of the latter and may perhaps be overcome
in a future theory of quantum gravity.

In this respect the similarity of the above proposal of phase charges
to Anandan's conception of gauge fields as ``interference fields''
should perhaps be emphasized. For the particular case
of general relativity, Anandan (1979) 
has shown that the ``gravitational phase'' is
$\Delta \chi = \frac{mc}{\hbar} \int g_{\mu\nu} dx^{\mu\nu}$,
that is the spacetime distance measured in Compton wavelengths and,
hence, $m$ being the inertial mass!
The same is true for neutron interferometry, the famous COW experiments.
Here the phase shift is actually proportional to the product of
inertial and gravitational mass, since the gravitational potential
includes the latter (cf. Greenberger 1983). 

A possible objection against the phase charge proposal is that,
in order to experimentally realize non-trivial bundles with
flat connections, a region with a non-flat connection (i.e.\
a real field and, hence, field charges) must exist simultaneously.
This seems to show that we cannot observe $q^{(p)}$ independently
from $q^{(f)}$. But again: the two charges are of different origin
in the sense that we take, for instance, the electron's field
charge to realize the electric current in the solenoid,
but may perform the AB experiment with electrons, muons or tauons
to measure their phase charge.
Moreover, strictly speaking we cannot make an independent
measurement of the inertial mass either. We always have to make
the idealization of neglecting the gravitational masses of the
measuring devices.

The reader may finally ask what the world were looking like if the
proposed equivalence between phase and field charge would empirically
be violated. The astonishing answer is that this would
not change so much the phenomenology of our elementary particles
world, since the violation only becomes visible in experiments
where non-trivial holonomies with flat connections are involved.
Conceptually, however, such a violation would leave us
with a serious puzzle, since then---again---the explanatory
gap in the logical structure of the empirically so eminently 
successful gauge theories persists.
This gap is certainly not filled by the gauge `principle',
but may perhaps be construed as an equivalence principle
between phase and field charges.


\subsection*{Acknowledgements}

I would like to thank Harvey Brown and Alfred Pflug for helpful comments.


\section*{References}

\begin{description}

\item [] 
Anandan, J. (1979).
\newblock Interference, gravity and gauge fields.
\newblock {\em Nuovo Cimento}, 53 A(2):221--249.

\item[] 
Brading, K. and Brown, H.~R. (2003).
\newblock Observing gauge symmetry transformations?
\newblock {\em The British Journal for the Philosophy of Science}, fortcoming.

\item[] 
Eynck, T.~O., Lyre, H., and Rummell, N.~v. (2001).
\newblock A versus {B}! {T}opological nonseparability and the {A}haronov-{B}ohm effect.
\newblock (PITT-PHIL-SCI00000404).

\item[] 
Greenberger, D.~M. (1983).
\newblock The neutron interferometer as a device for illustrating the strange
  behaviour of quantum systems.
\newblock {\em Reviews of Modern Physics}, 55(4):875--905.

\item[] 
Lyre, H. (2000).
\newblock A generalized equivalence principle.
\newblock {\em International Journal of Modern Physics D}, 9(6):633--647.
\newblock (gr-qc/0004054).

\item [] 
't~Hooft, G. (1980).
\newblock Gauge theories of the forces between elementary particles.
\newblock {\em Scientific American}, 242(6):104--138.

\end{description}


\end{document}